\documentstyle[aps,twocolumn]{revtex}
\newcommand{\be}{\begin{equation}}
\newcommand{\ee}{\end{equation}}
\newcommand{\bea}{\begin{eqnarray}}
\newcommand{\eea}{\end{eqnarray}}
\newcommand{\OP}{\vec{\psi}}

\newcommand{\M}{\vec{m}}

\begin{document}

\title{Ordering kinetics of defect structures}
\author{Gene F. Mazenko and Robert A. Wickham }
\address{The James Franck Institute and the Department of Physics \\
         The University of Chicago \\
         Chicago, Illinois 60637 }
\date{\today}
\maketitle
%
%  ABSTRACT
%
\begin{abstract}
We show how the continuity equations expressing conservation
of topological point or string defect charge can be used to determine the
order-parameter correlation function for the phase-ordering kinetics of the
$O(n)$ model in the special case where the order parameter is constrained
to be near a defect core. In this regime we find a self-consistent solution by
assuming the order parameter is Gaussian.
The resulting linear equation for the order-parameter correlation
function has as its solution the Ohta-Jasnow-Kawasaki form.
\end{abstract}
\draft
In this letter  we show that the phase-ordering kinetics of defects in the
$O(n)$ model can be self-consistently evaluated assuming the order parameter
is a Gaussian field near the defect cores.  This, in turn, clarifies the
nature of recent determinations of defect correlation functions for the
$O(n)$ model \cite{LIU92b,MAZENKO97a,MAZENKO97b}. We consider systems with
either point defects ($n=d$, where $d$ is the spatial dimensionality) or
string
defects ($n=d-1$). In addition to their importance in condensed matter, these
systems are also relevant to problems in cosmological structure formation
\cite{REVIEW}.
Our approach is based on the recently derived
\cite{MAZENKO97c} continuity equation expressing conservation of topological
defect charge and differs from previous calculations which
 make more direct use of the
time-dependent Ginzburg-Landau (TDGL)
equation for the non-conserved order parameter
\cite{MAZENKO90,LIU92a}. The continuity equation for systems supporting string
 defects
has not appeared previously and is established here. These continuity
equations are used to find the equation satisfied by the order-parameter
correlation function where the order parameter is restricted to be near a
defect core. Since it is easy to confuse the usual order-parameter correlation
function with the correlation function for the order
parameter constrained to be near a defect core, we will, for
reasons which will become clear below, refer to this constrained
quantity as the  auxiliary field correlation function $f$.
We obtain a linear
equation for $f$ which has as
its solution the Ohta-Jasnow-Kawasaki (OJK) \cite{OJK} form.
Since the equation is linear, the Gaussian assumption is consistent.

Most of the focus in phase-ordering kinetics has been on developing theories
for the order-parameter correlation function.
The belief has developed that we have a fairly
good understanding of how to calculate this quantity in the late-time
scaling regime. Two theories have evolved which map the order parameter field
onto an auxiliary field and invoke the so-called Gaussian closure
approximation.  We shall refer to the correlation function for this
auxiliary field as $f_{OP}$.
The OJK method, with elaborations by others \cite{OONO88,BRAY93},
leads to a simple form for the auxiliary field correlation function.
The second approach, originated by Mazenko \cite{MAZENKO90},
determines $f_{OP}$ self-consistently
through the solution of a nonlinear eigenvalue problem.
A key point we make here is that it is not necessary that $f=f_{OP}$.

Since the order-parameter correlation function is a rather
structureless quantity which does not give a great deal of direct
information about the underlying disordering agents, Liu and Mazenko
\cite{LIU92b}  examined the correlations between the defects
themselves. The key element in this
work was, following Halperin \cite{HALPERIN81}, to identify the
positions of the defects with the zeros of the order-parameter field
which could, in turn, be mapped onto the zeros of a Gaussian auxiliary
field. The point defect charge density correlation function $G_{\rho}$
was determined using this technique in terms of $f$, assuming that $f=f_{OP}$.
For large defect separations, $G_{\rho}$ was found to agree
with numerical simulations \cite{MONDELLO90} and experiments \cite{NAGAYA95}
relevant to the case $n=2$.
However, for $n=2$ the theory produced an unphysical
divergence in $G_{\rho}$ at short-scaled distances $x$ due to a
non-analytic piece in auxiliary field correlation function $f_{OP}$
at small $x$.
This non-analytic piece occurs in the most elementary self-consistent
theory for $f_{OP}$ \cite{LIU92a} but Mazenko and Wickham
\cite{MAZENKO97a} have since shown
that one can construct the theory so that $f_{OP}$ is analytic in $x$.
This development highlights the difference between the order-parameter
correlation function, whose short distance non-analyticities are essential
and lead to the observed generalized Porod's law \cite{POROD} for the
structure factor, and $f$  which must be
smooth in order to have sensible theories of defect position and
velocity \cite{MAZENKO97d} correlations. The $f$ obtained from OJK is
smooth.

The approach we take in this paper is distinct and independent
from that used by OJK or that developed in \cite{MAZENKO90}.
 The technique used
 here is directly based on considerations of topological charge conservation
and is less directly based on the TDGL model.
 This leads to the important
difference that we treat quantities that are
constrained to be evaluated at the defect core. Although we recover the
OJK result we obtain information on the regime of applicability of this
result.

The system studied here has a defect dynamics generated by the
TDGL
model for a non-conserved $n$-component vector order
parameter $\OP (\vec{r},t)$:
%
% TDGL EQUATION
%
\be
\label{EQ:TDGL}
\frac{\partial \vec{\psi}}{\partial t}=\vec{K}\equiv
-\Gamma \frac{\delta F}{\delta \vec{\psi}}
\ee
where $\Gamma $ is a kinetic coefficient,  $F$ is a Ginzburg-Landau
effective free energy assumed to be of the form
\be
F=\int ~d^{d}r \biggl( \frac{c}{2}(\nabla \vec{\psi})^{2}
+V(|\vec{\psi}|)\biggr).
\ee
The coefficient $c$ is positive and the potential
$V$ is assumed to be of the $O(n)$-symmetric,
degenerate double-well form. We assume that the quench is from an initial
high-temperature disordered state to zero temperature so the usual
noise term in (\ref{EQ:TDGL}) is set to zero.

In previous work \cite{LIU92a}
on the order-parameter correlation function progress was
made by mapping the order parameter $\vec{\psi}$ onto an auxiliary field
$\vec{m}$, with the requirement that {\em away} from the defect cores
%
% HARD SPIN
%
\be
\label{EQ:HS}
\vec{\psi}=\psi_{0}\hat{m}
\ee
where $\psi_{0}$ is the magnitude of $\OP$ in the ordered phase.
Physically, we  interpret $\M$ to be the position relative to the
nearest defect and expect that {\em near} the defect cores
%
% NEAR THE CORE
%
\be
\label{EQ:NC}
\vec{\psi}=a\vec{m}+b \vec{m} (\vec{m})^{2}   + \dots
\ee
where the coefficients $a$ and $b$ depend on the details of the potential $V$.
Equations (\ref{EQ:HS}) and (\ref{EQ:NC}) represent topological
charge $\pm 1$ defects,
which have the lowest energy and dominate the late-time regime \cite{HIGHER}.
In the theory for order-parameter correlations \cite{LIU92a} property
(\ref{EQ:HS}) is crucial, whereas, in the theory of defect motion presented
here property (\ref{EQ:NC}) is relevant since we always work near the
defect cores.

In the simplest models \cite{MAZENKO90,LIU92a}, considered here,
the auxiliary field $\M$ is assumed to
be a Gaussian field with a normalized correlation function $f$ defined as
%
% F CORRELATION FUNCTION
%
\be
\delta_{\mu \nu} f(12) =
\frac{\langle m_{\mu} (1) m_{\nu} (2) \rangle}{\sqrt{S_{0}(1) S_{0}(2)}}
\ee
with $\delta_{\mu \nu} S_{0}(1) = \langle m_{\mu} (1) m_{\nu} (1) \rangle$.
Here we use the shorthand
$m_{\mu} (1) = m_{\mu} ( \vec{r}_{1},t_{1})$.
It is well-established that for late times $t$
following the quench the dynamics obey scaling and the system
can be described in terms of a single growing length $L(t)$,
which is characteristic of the spacing between defects. For a
non-conserved order-parameter $L(t) \sim t^{1/2}$ at late times.
Since we interpret $|\M|$ to be the distance to the nearest defect we
expect $|\M| \sim L$. At equal times ($t_{1}=t_{2}=t$) in the scaling regime
the auxiliary field correlation function can be written solely in terms
of the scaled length $x=|\vec{r}_{2} - \vec{r}_{1}|/L(t)$.
Hence  $f(12) = f(x)$.

The emphasis in this letter is on defect densities like the charge
density for point defects, given in terms of the order parameter by
\cite{HALPERIN81}
%
% POINT CHARGE DENSITY
%
\be
\rho =\delta(\OP){\cal D}
\label{eq:3.1}
\ee
where the Jacobian associated with the change of variables
from the set of defect positions to the field $\vec{\psi}$ is defined by
%
% JACOBIAN
%
\be
\label{EQ:JACOBIAN}
{\cal D}=\frac{1}{n!}\epsilon_{\mu_{1}\mu_{2} \dots \mu_{n}}
\epsilon_{\nu_{1}\nu_{2} \dots \nu_{n}}
\nabla_{\mu_{1}}\psi_{\nu_{1}}
\nabla_{\mu_{2}}\psi_{\nu_{2}} \dots
\nabla_{\mu_{n}}\psi_{\nu_{n}}.
\ee
$\epsilon_{\mu_{1}\mu_{2}...\mu_{n}}$ is the
$n$-dimensional fully anti-symmetric tensor and
summation over repeated indices in (\ref{EQ:JACOBIAN}) is implied.

It was shown in a direct manner in \cite{MAZENKO97c} that $\rho$ satisfies the
continuity equation for topological charge
\be
\label{EQ:PTCON}
\frac{\partial\rho}{\partial t}=\nabla_{\alpha}[\delta
(\vec{\psi})J_{\alpha}
^{(K)}]
\ee
where the current $J_{\alpha}^{(K)}$
is defined as
%
% POINT DEFCT CURRENT
%
\be
\label{EQ:POINTCURRENT}
J_{\alpha}^{(K)}
=\frac{1}{(n-1)!}\epsilon_{\alpha\mu_{2} \dots \mu_{n}}
\epsilon_{\nu_{1}\nu_{2} \dots \nu_{n}}
K_{\nu_{1}} \nabla_{\mu_{2}}\psi_{\nu_{2}} \dots
\nabla_{\mu_{n}}\psi_{\nu_{n}}.
\ee
The derivation of  (\ref{EQ:PTCON}) is independent of the
details of the TDGL model, except that equation (\ref{EQ:TDGL})
is first order in time.
Since $\vec{J}^{(K)}$ is multiplied by the defect-locating
$\delta$-function we can replace $\vec{K}$ in $\vec{J}^{(K)}$ by the
part of $\vec{K}$ which does not vanish as $\vec{\psi}\rightarrow 0$.
For a non-conserved order parameter this means that we can set $\vec{K} =
\Gamma c \nabla^{2} \OP$ in (\ref{EQ:PTCON}).
Equation (\ref{EQ:PTCON}) is in the standard form of a continuity equation,
allowing us to identify the vortex velocity field as
\be
v_{\alpha}=-\frac{J_{\alpha}^{(K)}}{{\cal D}}
\ee
where it is assumed that the velocity field is used inside expressions
multiplied by the vortex-locating  $\delta$-function.

For string defects the charge density is a vector
given by \cite{HALPERIN81}
\be
\rho_{\alpha}
=\delta(\vec{ \psi})\omega_{\alpha}
\ee
with
\be
\omega_{\alpha}
=\frac{1}{n!}\epsilon_{\alpha\mu_{1}\mu_{2} \dots \mu_{n}}
\epsilon_{\nu_{1}\nu_{2} \dots \nu_{n}}
\nabla_{\mu_{1}}\psi_{\nu_{1}}
\nabla_{\mu_{2}}\psi_{\nu_{2}} \dots
\nabla_{\mu_{n}}\psi_{\nu_{n}}.
\ee
As in \cite{MAZENKO97c} one can obtain the continuity equation
satisfied by
$\rho_{\alpha}$ by combining the two identities
\be
\dot{\omega}_{\alpha}=\nabla_{\beta}J_{\alpha\beta}^{(K)}
\ee
and
\be
K_{\gamma}\omega_{\alpha}=J_{\alpha\beta}^{(K)}
\nabla_{\beta}\psi_{\gamma}
\ee
to obtain
\be
\frac{\partial\rho_{\alpha}}{\partial t}=\nabla_{\beta}[\delta
(\vec{\psi})J_{\alpha\beta}
^{(K)}].
\ee
The string defect current tensor $J_{\alpha \beta}^{(K)}$ is defined as
\be
J_{\alpha\beta}^{(K)}
=\frac{1}{(n-1)!}\epsilon_{\alpha\beta\mu_{2} \dots \mu_{n}}
\epsilon_{\nu_{1}\nu_{2} \dots \nu_{n}}
K_{\nu_{1}}
\nabla_{\mu_{2}}\psi_{\nu_{2}} \dots
\nabla_{\mu_{n}}\psi_{\nu_{n}}.
\ee
For the important case of $n=2$, $d=3$ we can write
\be
\label{EQ:J23}
J_{\alpha\beta}^{(K)}= v_{\alpha} \omega_{\beta} - v_{\beta} \omega_{\alpha}
\ee
and identify the string velocity as
\be
\label{EQ:VELSTR}
\vec{v}=\frac{1}{\omega^{2}}\left(\vec{\omega}\times\vec{g}\right)
\ee
where $\vec{g}=\epsilon_{\nu_{1}\nu_{2}}K_{\nu_{1}}
\vec{\nabla}\psi_{\nu_{2}}$ \cite{DZALOSHINSKI81}.

We use the continuity equation for topological charge
 to determine the auxiliary field correlation function.
We begin by examining point defects and require that the exact equation
\bea
\lefteqn{
\frac{\partial}{\partial t} \langle \rho(1)\rho(2) \rangle =
\nabla_{\beta}^{(1)} \langle \delta[ \vec{\psi} (1)]
J_{\beta}^{(K)}(1)\rho(2) \rangle
} \nonumber \hspace{2 in} \\ & &
\hspace{-1 in}
+\nabla_{\beta}^{(2)} \langle
\rho(1)\delta[ \vec{\psi} (2) ] J_{\beta}^{(K)}(2) \rangle
\label{EQ:OFMOT}
\eea
be satisfied at equal times.
The presence of the $\delta$-functions in (\ref{EQ:OFMOT}) enables us to
use relation (\ref{EQ:NC}) to replace $\vec{\psi}$ with
the Gaussian auxiliary field $\vec{m}$ in (\ref{EQ:OFMOT}).
The left-hand side of (\ref{EQ:OFMOT}) involves the     point defect
charge density correlation function    defined by
%
% VORTEX CHARGE DENSITY CORRELATION FUNCTION
%
\be
G_{\rho} (12)= \langle \rho(1) \rho(2) \rangle.
\ee
As shown in \cite{LIU92b}, $G_{\rho}$ factors into a product of
Gaussian averages which can be evaluated using standard methods.
In the scaling regime $G_{\rho}$ has the form
\be
G_{\rho}(12)=\frac{g(x)}{L^{2n}}
\ee
with $g(x)$ given by
\be
\label{EQ:DEFECTDEFECTCOR}
g (x) = n! \left[ \frac{h(x)}{x} \right]^{n-1}
\frac{\partial h(x)}{\partial x}.
\ee
We define
\be
\label{EQ:HDEF}
h = - \frac{\gamma f'}{2 \pi}
\ee
with $\gamma = 1/\sqrt{1-f^2}$.
It is therefore clear that the left-hand side of (\ref{EQ:OFMOT}) is a
complicated non-linear function of $f$ and its derivatives.
The right-hand side of (\ref{EQ:OFMOT}) involves the evaluation of
\be
N_{\beta}(12)= \langle \delta[\vec{m}(1)]{\cal D}(1)\delta[\vec{m}(2)]
J_{\beta}^{(K)}(2) \rangle
\ee
which, like the evaluation of $G_{\rho}$, factors
into products of averages over the $n$ separate components of
$\M$.  We easily find
\be
N_{\beta}(12)=n!\Gamma c B(A)^{n-1}	\hat{x}_{\beta}
\ee
where, in the scaling regime, $A$ is given by
\be
\label{EQ:A}
A= \frac{1}{L^{2}} \frac{h}{x}
\ee
and $B$ is
\be
\label{EQ:B}
B=\frac{\tilde{B}}{L^{3}}
\ee
with
\be
\label{EQ:TB}
2\pi \tilde{B}=\frac{d}{dx}\left[\gamma\left(\nabla^{2}f
+\frac{ n S^{(2)}}{\sigma} f\right)\right].
\ee
We have defined $S^{(2)} = \sum_{\alpha\beta}
\langle [\nabla_{\alpha} m_{\beta} ]^{2}
\rangle/n^{2}$ and
$\sigma = S_{0}/L^{2}$, which are both constants at late times.

Having compiled these results, it is  easy to see that
(\ref{EQ:OFMOT}) reduces, in the scaling regime, to
%
% SCALING FORM FOR POINT
%
\be
\label{EQ:INTERMEDIATE}
\mu\frac{d}{dx}\left[h^{n}+xh^{n-1}h'\right]
=\frac{d}{dx}\left[\tilde{B}h^{n-1}\right]
\ee
where we have defined the constant
\be
\mu=\frac{L\dot{L}}{2\Gamma c}.
\ee
Equation (\ref{EQ:INTERMEDIATE}) can be integrated to give
\be
\label{EQ:INTER2}
\mu\frac{d}{dx}(xh)=\tilde{B}
\ee
with the integration constant determined to be zero by the condition
that $f$ and $h$ vanish as $x\rightarrow \infty$.
Equation (\ref{EQ:TB}) shows us that $\tilde{B}$ is the derivative of a
quantity which vanishes at $x\rightarrow \infty$. Thus we can
integrate (\ref{EQ:INTER2}) yet again, and use (\ref{EQ:HDEF}), to obtain
the remarkable final result
\be
\label{EQ:FINAL}
-\mu x f'=\nabla^{2}f+\frac{nS^{(2)}}{\sigma} f
\ee
which is linear in $f$.

For string defects, the analogous calculation involves
evaluating the averages in
\bea
\lefteqn{
\frac{\partial}{\partial t} \langle \rho_{\alpha}(1)\rho_{\beta}(2) \rangle=
\nabla_{\gamma}^{(1)} \langle \delta[\vec{\psi} (1)] J_{\alpha\gamma}^{(K)}(1)
\rho_{\beta}(2) \rangle
} \hspace{2 in} \nonumber \\ & &
\hspace{-1 in}
 +
\nabla_{\gamma}^{(2)} \langle \rho_{\alpha}(1)\delta[\vec{\psi} (2)]
J_{\beta\gamma}^{(K)}(2) \rangle .
\label{EQ:STRMOT}
\eea
The string charge density correlation function
\be
G_{\alpha\beta}(12)= \langle \rho_{\alpha}(1)\rho_{\beta}(2) \rangle
\ee
was first worked out by Liu and Mazenko \cite{LIU92b} with the scaling result
\be
\label{EQ:STRCOR}
G_{\alpha\beta}(12)=\frac{1}{L^{2n}} \left[ G_{T}(x)
 \left(\delta_{\alpha\beta}- \hat{x}_{\alpha} \hat{x}_{\beta} \right)
+G_{L}(x) \hat{x}_{\alpha} \hat{x}_{\beta} \right]
\ee
where the transverse function is
\be
G_{T}(x)=n!\left(\frac{h}{x}\right)^{n-1}\frac{\partial h}{\partial x}
\ee
and the longitudinal function is
\be
G_{L}(x)=n!\left(\frac{h}{x}\right)^{n}.
\ee
$h$ is defined in (\ref{EQ:HDEF}).
We evaluate the average appearing on the right-hand side of (\ref{EQ:STRMOT})
using the same techniques as used in the point defect case
and easily obtain
%
% AVERAGE INVOLVING THE STRING CURRENT
%
\be
\label{EQ:CURAVG}
 \langle \delta[\M(1)] J_{\alpha\gamma}^{(K)}(1)\rho_{\beta}(2) \rangle
=n!\Gamma c B(A)^{n-1}
\left[\hat{x}_{\alpha}\delta_{\beta\gamma}
-\hat{x}_{\gamma}\delta_{\alpha\beta}\right]
\ee
where $A$ is defined in (\ref{EQ:A}) and $B$ in (\ref{EQ:B}).
Substitution of the results (\ref{EQ:STRCOR}) and (\ref{EQ:CURAVG}) into
(\ref{EQ:STRMOT}) leads to separate equations for
the longitudinal and transverse components.
Both equations reduce to (\ref{EQ:FINAL}). Thus the same linear
equation determines $f$ for both point and string defects.

The solution of (\ref{EQ:FINAL})  is of the OJK form
%
% OJK FORM
%
\be
f(x)=\exp{-\frac{\mu}{2} x^{2}}
\ee
where we have used the relation
%
% RELATION BETWEEN LENGTHS
%
\be
\frac{n S^{(2)}}{\sigma} \equiv  \left(-\nabla^{2}f\right)|_{x=0}= n \mu.
\ee
The constant $\mu$ is fixed through a choice of the length scale $L$, since
$L = \sqrt{4 \Gamma c \mu} t^{1/2}$.
We are therefore lead directly, without any approximation except that
the order parameter field can be treated as Gaussian near its zeros,
to a self-consistent result.

It should be noted that at leading order in a
systematic large-$N$ approximation scheme for a colour index $N$
Bray and Humayun \cite{BRAY93} were able to recover the previously
{\em ad hoc}
Oono-Puri \cite{OONO88} extension of OJK which, in turn, gives {\em exactly}
 (\ref{EQ:FINAL}).  In the development here, equation
 (\ref{EQ:FINAL}) arises without the need for any
such approximations.

This work highlights the fact that the original OJK result
 is implicitly derived
near the defect cores \cite{OJK}.
It also emphasizes the distinction between theories of the
order parameter correlation function, such as in Ref. \cite{MAZENKO90},
and theories concerning defect correlation functions,
where it appears one can self-consistently use the Gaussian closure
approximation if one uses the OJK form for the
auxiliary field correlation function.
The theory of order-parameter correlations does
a superior job determining the non-equilibrium exponent governing the
 decay of two-time autocorrelation functions \CITE{LIU92a,LIU91}, but the
use of the theory presented here for defect correlations,
with its smooth OJK-like auxiliary field,
avoids the difficulties due to non-analyticities at small $x$.
%
% ACKNOWLEDGMENTS
%
\acknowledgements

This work was supported primarily by the MRSEC Program of the National Science
Foundation under Award Number DMR-9400379.
%
% REFERENCES
%


\begin{thebibliography}{99}
\bibitem{HIGHER}     More generally, near the defect core the order
                     parameter vanishes as $m^{q}$ where $q$ is
                     the magnitude of the topological charge of
                     the defect.
\end{thebibliography}
\end{document}